\documentclass[conference]{IEEEtran}
\usepackage{amsmath,amsfonts}
\usepackage{algorithmic}
\usepackage{algorithm}
\usepackage{xcolor}
\usepackage{amssymb}
\usepackage{array}
\usepackage{graphicx}
\usepackage{multirow}
\usepackage{makecell}
\usepackage{babel}
\usepackage{cite}
\usepackage{gensymb}
\usepackage{mathtools}
\usepackage[bbgreekl]{mathbbol}
\usepackage[thmmarks, amsmath]{ntheorem}

\DeclareSymbolFontAlphabet{\mathbb}{AMSb}
\DeclareSymbolFontAlphabet{\mathbbl}{bbold}

\DeclareMathOperator{\diff}{d}
\DeclareMathOperator{\diag}{diag}

\newtheorem{theorem}{Theorem}
\newcommand{\mc}{\mathcal}
\newcommand{\ddt}{\tfrac{\diff}{\diff \!t}}
\newcommand{\dc}{{\text{dc}}}

\newcommand{\g}{{\text{g}}}
\newcommand{\gsc}{{\text{gsc}}}
\newcommand{\msc}{{\text{msc}}}
\newcommand{\del}{{\text{del}}}
\newcommand{\mpp}{{\text{mpp}}}
\newcommand{\w}{{\text{w}}}

\newcommand{\wt}{{\text{wt}}}
\newcommand{\pmsg}{{\text{pmsg}}}
\newcommand{\refe}{{\text{ref}}}

\usepackage{url}

\usepackage[]{fancyhdr} %
\newcommand{\changefont}{\fontsize{9}{9}\selectfont}
\IEEEoverridecommandlockouts
\begin{document}

\title{Unified Grid-Forming Control of PMSG Wind Turbines for Fast Frequency Response and MPPT}
\author{\IEEEauthorblockN{Xue Lyu, Dominic Gro\ss{}}
\IEEEauthorblockA{Department of Electrical and Computer Engineering\\University of Wisconsin Madison\\
Madison, United States} 
\and
\IEEEauthorblockN{Irina Suboti\'c}
\IEEEauthorblockA{Automatic Control Laboratory\\ ETH Z\"urich\\
Z\"urich, Switzerland}\thanks{This work was partially funded by the Swiss Federal Office of Energy under grant number SI/501707.} 
}

\maketitle
\thispagestyle{fancy}
\pagestyle{fancy}

\begin{abstract}
In this work we present a novel dual-port grid-forming control strategy, for permanent magnet synchronous generator wind turbines with back-to-back voltage source converters, that unifies the entire range of functions from maximum power point tracking (MPPT) to providing inertia and fast frequency response without explicit mode switching between grid-following and grid-forming control. The controls impose a well-defined AC voltage at the grid-side converter (GSC) and the machine-side converter (MSC) AC terminals and explicitly stabilize the DC-link capacitor voltage through both GSC and MSC. The wind turbine's kinetic energy storage and curtailment are adjusted through a combination of implicit rotor speed control and pitch angle control and directly determine the operating mode and level of grid support. Moreover, we provide analytical small-signal stability conditions for a simplified system and explicitly characterize the relationship between control gains, curtailment, and the wind turbines steady-state response. Finally, a detailed simulation study is used to validate the results and compared the proposed control with state-of-the-art MPPT control.
\end{abstract}

\IEEEpeerreviewmaketitle

\section{Introduction}
The dynamics of conventional bulk power systems are dominated by synchronous generators (SGs) that regulate the system voltage and frequency through a combination of the synchronous generators' inherent physical property (e.g., rotational inertia) and controls (e.g., governor, voltage regulator). In contrast, today's grid-following (GFL) converter-interfaced resources require a stable power system to operate, inherently rely on a phase-locked loop (PLL) for grid-synchronization, and ultimately jeopardize system stability \cite{WEB+15,MDH+18}. Therefore, to achieve significant integration of renewables, grid-forming (GFM) converters that do not require external voltage/frequency regulation but impose an AC voltage waveform with a well defined frequency and magnitude at their point of common coupling (PCC) are envisioned to be the cornerstone of future power systems \cite{MDH+18}.

The prevalent GFM control methods for DC/AC voltage source converters (VSC) can be broadly categorized into droop control \cite{chandorkar1993control}, virtual synchronous machines \cite{d2015virtual}, and virtual oscillator control \cite{johnson2013synchronization, gross2019effect}. All aforementioned methods impose an AC voltage with well-defined amplitude and frequency that vary in response to active and reactive power injection but assume a stable DC voltage (i.e., a fully controllable power source) and do not consider power source dynamics or limitations that can result in instability \cite{TGA+20}. 

Instead, in this work we leverage the recently proposed dual-port GFM control paradigm \cite{gross2021dual,SG2021} that maintains stability if either the AC or DC terminal voltage are stable or the AC or DC side power generation has sufficient flexibility. In this work, we apply the dual-port GFM paradigm to permanent magnet synchronous generator wind turbines with back-to-back voltage source converters to support functions ranging from maximum power point tracking to inertia support and fast frequency response in a single unified grid-forming controller.

Several system operators already require wind power plants to have active power control capabilities to contribute to grid frequency regulation \cite{morales2008advanced}. Notably, active power control and further ancillary services (e.g., inertia response) can be realized without additional energy storage systems by exploiting the inherent kinetic energy storage and controllability of wind turbines \cite{wang2015control, krpan2020dynamic}. Specifically, rotor speed control and pitch angle control can be used to operate below the maximum power point (MPP) and provide primary frequency control  \cite{aho2016active, zertek2012novel, wang2019fast}. In the context of wind turbine's with back-to-back VSCs, various GFL \cite{lyu2018coordinated,zertek2012novel, wang2019fast} and (partial) GFM \cite{NRELGFM,YCG+2021} control strategies have been proposed to enable an inertia response and/or primary frequency control capabilities. GFL implementations rely on PLL-based grid-side converter (GSC) control for (i) DC voltage stabilization, and (ii) to obtain frequency measurements that are used to modify the active power setpoint of the machine-side converter (MSC). In contrast, common GFM implementations aim to stabilize the DC voltage through the MSC control and use standard GFM controls on the GSC \cite[Sec. II-F]{YCG+2021}. Crucially, conventional controls require switching between controls to support the entire range of functions (i.e., from MPPT control to GFM frequency control) and, using GFM control, coordinating the curtailment, AC frequency, and DC voltage control can be a complex task.

In contrast, we propose a control strategy that always imposes an AC voltage with well-defined amplitude and frequency at the PCC, does not rely on a PLL, supports both MPPT and frequency response functions (e.g., inertia, fast frequency response) without explicit mode switching, and has relatively low complexity. In particular, the ancillary services provided (e.g., inertia, fast frequency response) are fully determined by the wind turbine operating point and resulting flexibility, e.g., at the MPP no primary frequency control is provided, while operating at a rotor speed beyond the MPP allows for an inertia response and primary frequency control. Using the proposed control strategy rotor speed and pitch control are used collaboratively and the achievable frequency response can be explicitly linked to the wind turbine dynamics and flexibility at the curtailed operating point. Specifically, we provide analytical small-signal stability conditions for a simplified test system and propose a method to select the key control parameters based on steady-state control specifications. Finally, a detailed simulation study is used to compared the proposed control to standard MPPT controls and illustrate that the proposed control (i) supports standard control functions (i.e., MPPT, inertia response, fast frequency response), and (ii) provides significant grid support.

\section{Wind turbine model}
In this work, we focus on wind turbines using a permanent magnet synchronous generator (PMSG) interfaced through back-to-back voltage source converters. This section briefly reviews the models used to design and analyze the proposed dual-port grid-forming controller. 

\subsection{Wind turbine aerodynamics}
\begin{figure}[t!!]
\centering
\includegraphics[width=8.5cm]{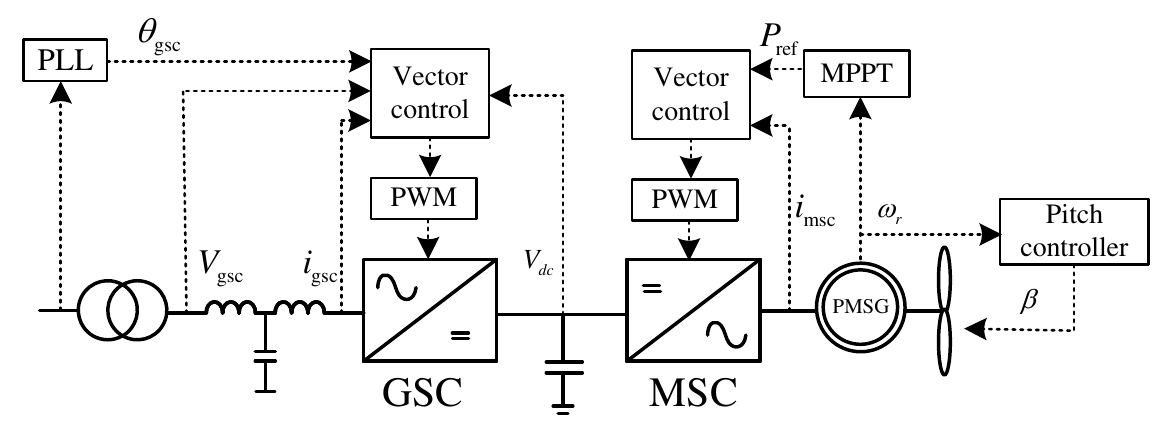}
\caption{Full-scale PMSG wind turbine with back-to-back voltage source converters, DC-link capacitor, and grid following MPPT control.}\label{fig:WT GFL control}
\vspace{1em}
\centering
\includegraphics[width=7cm]{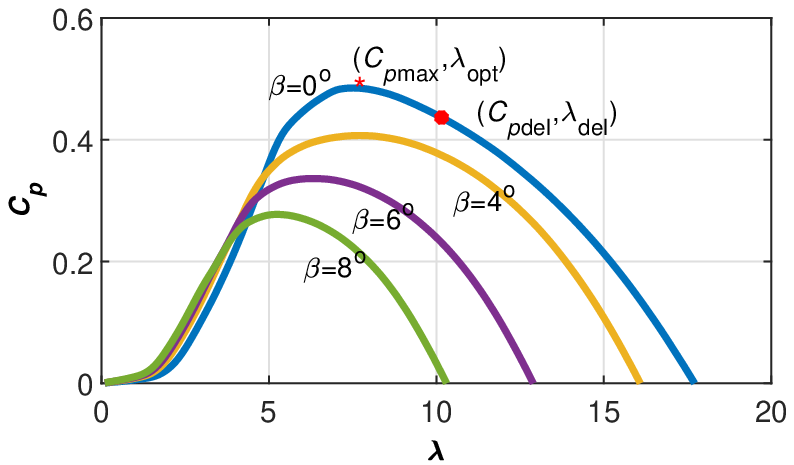}
\caption{Power coefficient $C_p$ of a $5~\mathrm{MW}$ wind turbine \cite{NREL5MW} as function of the pitch angle $\beta$
and tip speed ratio $\lambda$. $C^\mpp_p$ denotes the maximum power coefficient and $\lambda_\mpp$ denotes the corresponding tip speed ratio at $\beta=0$. Moreover, $C^\del_p$ and $\lambda_\del$ denotes an operating point with wind power generation curtailment achieved by increasing the rotor speed.\label{fig:Cp}}
\end{figure}

The mechanical power generated by a wind turbine is 
\begin{align}
P_\wt=\frac{1}{2} \rho \pi R^2 C_p (\lambda, \beta) v_w^3,
\end{align}
where $\rho \in \mathbb{R}_{>0}$ denotes the density of air, $R \in \mathbb{R}_{>0}$ denotes the rotor radius, $v_\w \in \mathbb{R}_{>0}$ denotes the wind speed. Moreover, $C_p: \mathbb{R}_{>0} \times \mathbb{R}_{>0}  \to \mathbb{R}_{>0}$ models the fraction of the wind power captured as function of the blade pitch angle $\beta \in \mathbb{R}_{\geq0}$
and the so-called tip speed ratio $\lambda = R \omega_r / v_\w$ with rotor speed $\omega_r \in \mathbb{R}_{>0}$. The power coefficient as function of the pitch angle $\beta \in \mathbb{R}_{>0}$
and the tip speed ratio $\lambda \in \mathbb{R}_{>0}$ is shown in Fig.~\ref{fig:Cp} and can, within limits, be adjusted through the rotor speed $\omega_r$ and blade pitch angle $\beta$.
\subsection{Power conversion}
The mechanical power $P_\wt$ generated by the wind turbine is converted to electrical power using an electric machine. For brevity of the presentation, we consider a single-mass model of the rotor and reduced-order synchronous machine model
\begin{subequations}\label{eq:machine.model}
\begin{align}
\ddt \theta_r &= \omega_r,\\
J_\wt \omega_r \ddt \omega_r &= P_\wt - P_\pmsg \label{eq:machine.model.omega.}. 
\end{align}	  
\end{subequations}
where $P_\pmsg \in \mathbb{R}$ denotes the active power flowing from the PMSG into the MSC, $\theta_r$ denotes the AC voltage phase angle of the PMSG, and $J_\wt \in \mathbb{R}_{>0}$ denotes the combined inertia of the rotor, blades, and PMSG.

Moreover, each DC/AC voltage source converter (VSC) modulates the DC-link capacitor voltage $v_\text{dc}$ into an AC voltage. For the purpose of control design, we consider a reduced-order averaged model for which the AC voltage phase angle (i.e., $\theta_\gsc$) and magnitude (i.e., $V_\gsc$ are control inputs that are either tracked by inner controls (see e.g., \cite{d2015virtual}) or directly modulated at the converter switching terminal. The only remaining dynamics are the DC-link capacitor charge dynamics:
\begin{align} \label{eq:converter.model}
C_\dc v_\dc  \ddt v_\dc &= - P_\msc - P_\gsc = P_\pmsg - P_\gsc, 
\end{align}
where $C_\dc \in \mathbb{R}_{>0}$ denotes the DC-link capacitance and $P_\gsc$ and $P_\msc = -P_\pmsg$  denote the active power flowing out of the grid-side converter (GSC) and motor-side converter (MSC).
\subsection{Power exchange}
Finally, for the purpose of control design and analysis, we model the active power $P_\pmsg$ flowing from the PMSG into the MSC and the active power $ P_\gsc$ flowing from the GSC to the system through the lossless AC power flow model
\begin{subequations}\label{eq:powerflow}
 \begin{align}
  P_\pmsg &= b_\msc \sin(\theta_r - \theta_\msc), \label{eq:mscpower}\\
  P_\gsc &= \sum\nolimits_{k=1}^{n_b} b_{\gsc,k} \sin(\theta_\gsc-\theta_k),\label{eq:gscpower}
 \end{align}
\end{subequations}
with susceptance $b_\msc$ that models the combined impedance of the MSC filters and PMSG stator. Similarly, $b_{\gsc,k}$ denotes the line/transformer susceptances and $\theta_k$ models the AC voltage phase angles at the $n_b \in \mathbb{N}$ buses in the AC power system.

\section{Review of GFL and GFM control strategies}
Control strategies for wind turbines broadly fall into two categories, grid-following and grid-forming that are complementary in the sense that grid-following control is needed to achieve maximum power point tracking, while grid-forming control is the preferred solution for providing grid support. In this section, we briefly review the two paradigms. A particular challenge is that (1) wind turbine's in future systems may have to switch between different controls depending on their operating point and functions provided to the grid (e.g., MPPT vs. primary frequency control), and (2) the dynamic interactions between heterogeneous grid-following and grid-forming controls can result in complex dynamics and adverse dynamic interactions.

\subsection{Grid-following MPPT control}
Today, wind turbines typically operate in MPPT mode to maximize wind power generation  (see Fig.~\ref{fig:WT GFL control}). To this end, the rotor speed is adjusted to operate at the optimal tip speed ratio $\lambda_\mpp$ (see Fig.~\ref{fig:Cp}). If the optimal rotor speed $\omega^{\mpp}_r = v_\w \lambda_\mpp / R$ exceeds the maximum rotor speed $\omega^{\max}_r$, the blade pitch angle is increased to reduce power generation and limit the rotor speed to $\omega^{\max}_r$. In grid-following control mode, the GSC and MSC utilize a vector AC current control in synchronous reference frame. The reference angle for the GSC is provided by a phase-locked loop, while the reference angle for the MSC is the machine rotor angle. Commonly, a proportional-integral (PI) controller for the DC voltage and reactive power controller are used to determine the current reference for the GSC. In contrast, an outer rotor speed or torque controller provides the MSC active current to control the tip speed ratio $\lambda$ to $\lambda_\mpp$ (see e.g., \cite{aho2016active}) and reactive current is controlled to zero.

\subsection{Grid-forming control}
Conceptually, any grid-forming control (e.g., \cite{chandorkar1993control,johnson2013synchronization,d2015virtual,gross2019effect}) can be used on the GSC as long as the DC voltage is controlled to its nominal value $v^\star_\dc$ through the MSC. For example, a DC voltage PI controller can be used to provide a current reference to an underlying MSC vector current controller to control $v_\dc$ to $v^\star_\dc$. However, if the MSC is overloaded or exceeds the reserves of the power source the DC-voltage will become unstable \cite{TGA+20}. While additional outer controls (see e.g. \cite{LCP20}) can be used to mitigate overload, coordinating the curtailment, AC frequency, DC voltage, wind turbine rotor speed and blade pitch control is a complex task.

\section{Dual-Port Grid-Forming Control of Wind Turbines}
Instead of relying on either the GSC or MSC to control the DC voltage, we apply the so-called dual-port GFM control paradigm to wind turbines. In particular, dual-port GFM control imposes a well defined AC voltage waveform on both GSC and MSC AC terminals and controls the DC voltage through both GSC and MSC. In this section, we will first discuss the curtailment strategy used to provide reserves, the GSC and MSC control, as well as the blade pitch angle control.

\subsection{Overview of the control strategy}
\begin{figure}
\centering
\includegraphics[width=1\columnwidth]{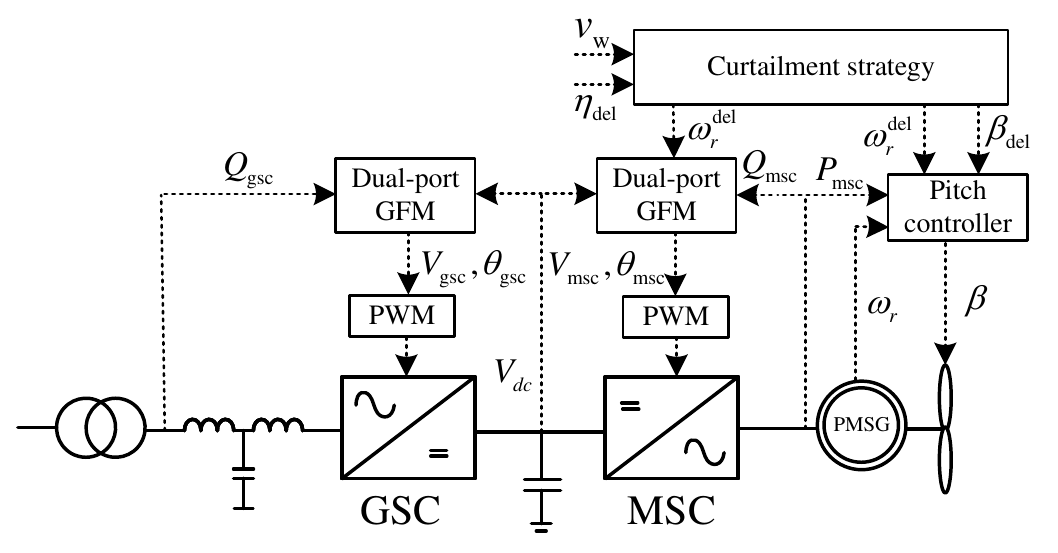}
\caption{{Overview of dual-port grid forming control for PMSG wind turbines.}\label{fig:WTGFM}}
\end{figure}
{Before proceeding with a detailed description of the controls for the different components of the PMSG wind turbine system we first discuss how the different components and controls interact. The overall system comprised of two back-to-back voltage source converters, a permanent magnet synchronous generator, a wind turbine, and their corresponding controls are shown in Fig.~\ref{fig:WTGFM}. The MSC and GSC are controlled by dual-port GFM controls \cite{gross2021dual} that control their AC voltage frequency deviation as a function of the DC-link voltage deviation (see Sec.~\ref{sec:GSCcont} and Sec.~\ref{sec:MSCcont}). The dual-port GFM controls induce GSC frequency dynamics that self-synchronize with the grid and MSG frequency dynamics that self-synchronize with the permanent magnet synchronous generator of the WT.  Therefore, after a load disturbance (i.e., a load step) in the grid, the steady state deviations of the GSC frequency $\omega_\gsc$, DC-link voltage $v_\dc$, and rotor speed $\omega_r$ will be proportional to each other. Thus, in MPPT mode, the DC-link voltage and the rotor speed, will be stabilized through the GSC AC side. In contrast, when grid-support is desired, the curtailment strategy (see Sec.~\ref{sec:curtailment}) computes in an operating point $(\omega_r^\del,\beta_\del)$ of the wind turbine at which a given fraction $\eta_\del \in \mathbb{R}_{[0,1]}$ of the maximum power is produced. At this curtailed operating point (see Fig.~\ref{fig:Cp}) and by design of the pitch controller (see Sec.~\ref{sec:pitch_controller}) a decrease in rotor speed $\omega_r$ results in an increase of the wind turbine power generation $P_\wt$. In other words, an increase in load results in a decrease of the DC-link voltage which in turn results in an increase of $P_\wt$.}

\subsection{Curtailment strategy}\label{sec:curtailment}
\begin{figure}
\centering
\includegraphics[width=5cm]{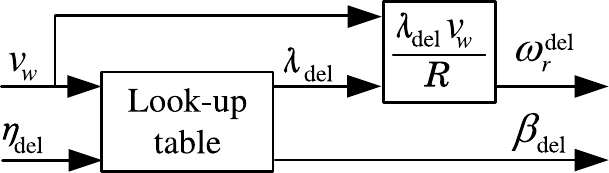}
\caption{{Online computation of the curtailed operating point  $(\beta_\del,\omega^\del_r)$. \label{fig:curtailment_strategy}}}
\end{figure}

In contrast to MPPT control, the wind turbine has to be operated below its maximum power point to provide standard grid-forming control functions such as inertia and primary frequency control response that require flexibility of the power generation to respond to changes in system load. Generally, curtailment of the wind power generation can be achieved by increasing the blade pitch angle and/or increasing the rotor speed $\omega_r$ beyond $\omega^{\mpp}_r$ \cite{aho2016active}. Moreover, depending on the wind speed, increasing the rotor speed $\omega_r$ beyond $\omega^{\mpp}_r$ can provide a significant amount of kinetic energy storage that can be used to provide grid support. Therefore, we prioritize curtailment through increased rotor speed over curtailment through increased pitch angles. To this end, we define the deloading parameter $\eta_\del \in \mathbb{R}_{[0,1]}$ and use $\lambda_\del \in \mathbb{R}_{\geq \lambda_\mpp}$ to denote the unique solution of 
\begin{align}\label{eq:speeddeload}
 C_p(\lambda_\del,0) = \eta_\del C_p(\lambda_\mpp,0)
\end{align}
that satisfies $\lambda_\del \geq \lambda_\mpp$.
If $\lambda_\del v_\w/R \leq \omega^{\max}_r$, we use the rotor speed  $\omega^\del_r \coloneqq \lambda_\del v_\w/R$ and $\beta_\del=0$. On the other hand, if $\lambda_\del v_\w/R > \omega^{\max}_r$, then $\omega^\del_r\coloneqq \omega^{\max}_r$ and $\beta_\del$ is given by the solution of 
\begin{align}\label{eq:pitchdeload}
 C_p(\tfrac{ R \omega^{\max}_r}{v_\w},\beta_\del) = \eta_\del C_p(\lambda_\mpp,0).
\end{align}
For real-time control, \eqref{eq:speeddeload} and \eqref{eq:pitchdeload} are solved offline for different wind speeds $v_\w$ and deloading parameters $\eta_\del$ {and the resulting values $\lambda_\del$ and $\beta_\del$ are stored in a lookup table. The lookup table is used to generate a curtailed operating point defined by $(\beta_\del,\omega^\del_r)$ as shown in Fig.~\ref{fig:curtailment_strategy}.}

\subsection{Dual-Port GFM Control of the Grid Side Converter}\label{sec:GSCcont}
The DC voltage $v_\dc$ of wind turbines with back-to-back converters reflects the active power balance between GSC and MSC (cf. \eqref{eq:converter.model}) and can vary within a limited range \cite{lyu2018coordinated}. This fact is used in the recently proposed dual-port GFM paradigm \cite{gross2021dual,SG2021} that leverages the duality between frequency and DC voltage as indicators of power imbalance \cite{CBB+2015, CGD17}. We obtain the GSC terminal voltage reference angle $\ddt \theta_\gsc = \omega_\gsc$ through the energy-balancing dual-port grid-forming control 
\begin{align} \label{eq:GSC voltage angle}
\omega_\gsc= \omega_0 + \left(\tfrac{K^\gsc_\theta+K^\gsc_d s}{T^\gsc_\dc s + 1}\right)(v_\dc-v_\dc^\star),
\end{align}
i.e., a realizable proportional-derivative (PD) control with proportional and derivative control gains $K^\gsc_\theta \in \mathbb{R}_{>0}$ and  $K^\gsc_d\in \mathbb{R}_{>0}$, and filter time constant $T^\gsc_\dc$ used to suppress switching harmonics from the DC voltage measurement. In other words, if the DC deviates from its nominal value, the GSC frequency is adjusted to change the GSC phase angle and adjust the GSC power injection \eqref{eq:gscpower}. This approach translates the signals indicating DC power imbalances (i.e., DC voltage) and AC power imbalances (i.e., frequency). In particular, after a load increase the DC voltage decreases resulting in a frequency drop. If the DC voltage is stabilized by the MSC, the GSC frequency is implicitly stabilized and grid support is provided. However, if the MSC does not stabilize the DC voltage, \eqref{eq:GSC voltage angle} results in DC voltage control through the GSC AC terminal.

\begin{figure}
\centering
\includegraphics[width=0.7\columnwidth]{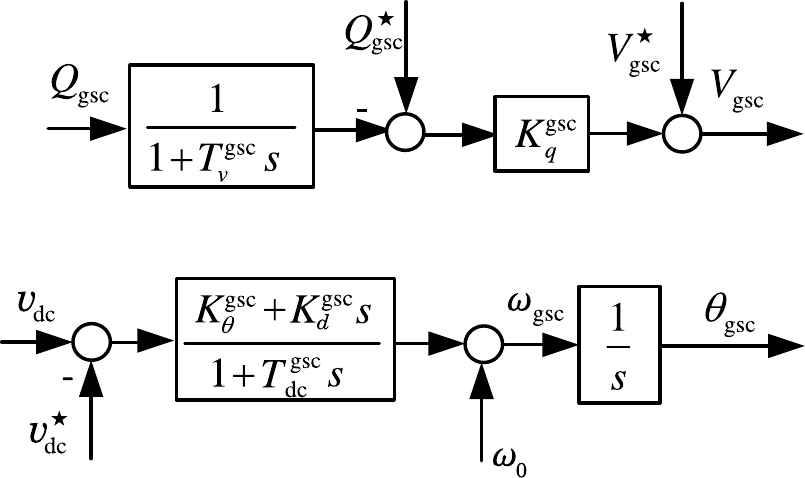}
\caption{Energy-balancing dual-port GFM control of the GSC.}\label{fig:GSC_controller}
\end{figure}

Finally, the deviation of the GSC AC voltage magnitude $V_\gsc$ from its nominal value $V_\gsc^\star$ is determined by standard $Q-V$ droop control \cite{d2015virtual}
\begin{align} 
V_\gsc=V_\gsc^\star + K^\gsc_q (Q_\gsc^\star-\tfrac{1}{T^\gsc_v s + 1}Q_\gsc),
\end{align}
with the reactive power set-point $Q_\gsc^\star$ and measured reactive power $Q_\gsc$. Moreover, $T^\gsc_v$ denotes  the reactive power low pass filter time constant, and $K^\gsc_q$ is the reactive power droop gain gain. A block diagram of the GSC control strategy is shown in Fig.~\ref{fig:GSC_controller}. 

\subsection{Dual-Port Grid-Forming Control of the Machine Side Converter}\label{sec:MSCcont}
\begin{figure}[t!]
\centering
\includegraphics[width=0.7\columnwidth]{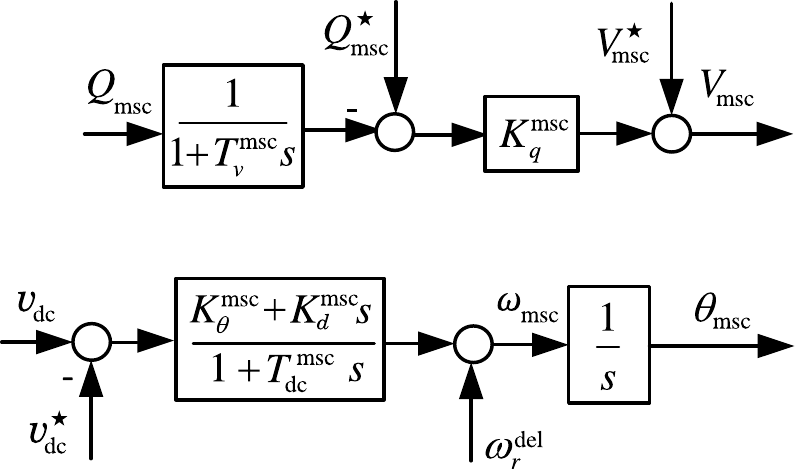}
\caption{{Energy-balancing dual-port GFM control of the MSC.}}\label{fig:MSC_controller}
\end{figure}

Both the GSC and MSC use the same dual-port grid-forming controller, i.e., 
\begin{subequations}
 \begin{align} 
\omega_\msc= \omega^\del_r + \left(\tfrac{K^\msc_\theta+K^\msc_d s}{T^\msc_\dc s + 1}\right) (v_\dc-v_\dc^\star),\label{eq:MSC voltage frequency}\\
V_\msc=V_\msc^\star + K^\gsc_q (Q^\star_\msc -\tfrac{1}{T^\msc_v s + 1}Q_\msc).
\end{align}
\end{subequations}
Here $\omega^\del_r \in \mathbb{R}_{\geq \omega^\mpp_r}$ denotes the rotor speed at the curtailed operating point, $V_\msc^\star$ the nominal MSC AC voltage magnitude, $Q_\msc^\star$ denotes the reactive power set-point and is typically set to zero to reduce machine side circulating currents and losses, and $Q_\msc$ denotes the measured reactive power on the machine side. Moreover, $T^\msc_\dc$ is a low pass filter time constant that can be used to suppress switching harmonics from the DC voltage measurement and  reduce the mechanical stress on the wind turbine. The block diagram of the dual-port GFM control used for the MSC is shown in Fig. \ref{fig:MSC_controller}.

In contrast to conventional control strategies that control the wind turbine's electric power generation through current control and an active power reference generated from a MPPT curve (grid-following MPPT) or DC-voltage controller (grid-forming), in this work, the wind turbine's electric power generation is controlled through the deviation of the MSC AC frequency $\omega_\msc$ from the nominal operating frequency
\begin{align}
 \omega^\del_r= \frac{\lambda_\del v_\w}{R} \geq  \omega^\mpp_r
\end{align}
In particular, $\eta_\del=1$ results in approximate MPPT control with $\lambda_\del = \lambda_\mpp$ and $\omega^\del_r=\omega^\mpp_R$. In this case, an increase in load (i.e., reduction in grid-frequency) results in a drop in DC voltage and rotor speed. However, at the MPP the sensitivity of the wind power generation to the rotor speed is approximately zero and only an inertia response is provided. In contrast, $\eta_\del \in \mathbb{R}_{[0,1)}$ results in curtailment and $\lambda_\del > \lambda_\mpp$ and $\omega^\star_r > \omega^\mpp_r$. Consequently, a reduction in DC voltage results in a drop in MSC frequency $\omega_\msc$, an increased power injection by the PMSG, and increased wind power generation (cf. Fig.~\ref{fig:Cp}). 
This results in an inertia and primary frequency response that stabilizes the DC voltage and GSC frequency. However, different control gains may be required for GSC and MSC, e.g., to achieve a specified frequency droop percentage (see Sec.~\ref{sec:gainsselection}).
 
\subsection{Blade Pitch Controller}\label{sec:pitch_controller}
\begin{figure}[h!]
\centering
\includegraphics[width=8cm]{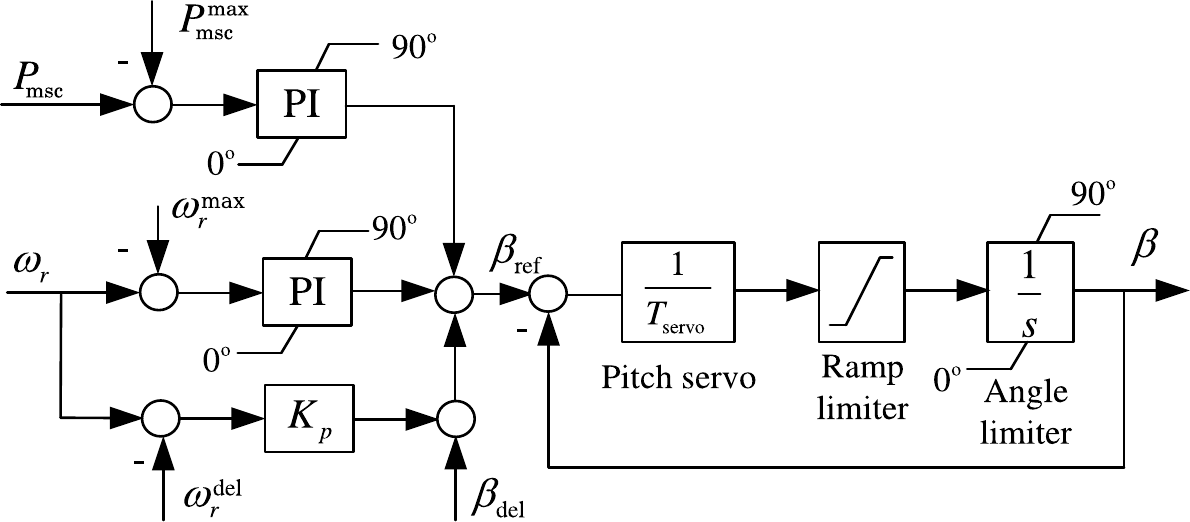}
\caption{Pitch controller for dual-port GFM control and pitch servo model. Two PI controls are used to limit the rotor speed and wind turbine power generation to their maximum values. \label{fig:pitch_controller}}
\end{figure}

Depending on the wind speed $v_\w$ and curtailment $\eta_\del$, both rotor speed and pitch angle control are used to curtail the wind power generation (see Sec.~\ref{sec:curtailment}). In particular, once the rotor speed setpoint $\omega^\del_r$ is equal to the maximum rotor speed $\omega^{\max}_r$, the proportional pitch controller 
\begin{align} \label{eq:beta_ref}
\beta_\refe=\beta_\del+K_p(\omega_r-\omega^\del_r)
\end{align}
with pitch angle setpoint $\beta_\del$ and gain $K_p \in \mathbb{R}_{\geq 0}$ is used to respond to power imbalances. The gain $K_p \in \mathbb{R}_{\geq 0}$ is zero if $\beta_\del=0$ and positive if $\beta_\del>0$. Moreover, the rotor speed and wind power generation need to be limited to account for the mechanical limits of the wind turbine and converter current limits. Therefore, two proportional-integral (PI) control loops are introduced that activate if $\omega_r\geq\omega^{\max}_r$ and $P_\msc \geq P^{\max}_\msc$ and reduce the pitch angle to protect the wind turbine and power converter as shown in Fig.~\ref{fig:pitch_controller}.

\section{Stability and Steady-State Analysis}\label{sec:analysis}
In this section we use a small-signal model to analyze dynamic stability and discuss the relationship between the control gains and the wind turbine's response to power imbalance. For brevity of the presentation, we will consider a simplified power system consisting of one synchronous generator  with turbine/governor system and one wind turbine operating in constant wind speed $v_\w$.

\subsection{Frequency dynamics and stability analysis}
To formalize the analysis, we approximate the deviation of the wind turbine power generation from the power generation $P^\del_\wt$ at the operating point $(\omega_r,\beta)\!=\!(\omega^\del_r,\beta_\del)$ by
\begin{subequations}\label{eq: WT power combine rotor & pitch}
\begin{align} 
P_\wt - P^\del_\wt &= K_{\omega_r}(\omega^\del_r-\omega_r)+K_\beta(\beta_\del-\beta),\\
&=(K_{\omega_r}+K_\beta k_p) (\omega^\del_r-\omega_r).\label{eq:WTpowerlin}
\end{align}
\end{subequations}
Here $K_{\omega_r}=\frac{\partial P_\wt}{\partial \omega_r} \rvert_{\scriptsize \omega_r=\omega^\del_r, \beta=\beta_\del}$ and 
$K_\beta=\frac{\partial P_\wt}{\partial \beta} \rvert_{\scriptsize \omega_r=\omega^\del_r, \beta=\beta_\del}$ denote the non-negative sensitivity of the mechanical power $P_\wt$ captured by the wind turbine with respect to a change in rotor speed or pitch angle. Moreover, the synchronous generator (SG) is modeled by the improved swing equation with first-order turbine governor model
\begin{subequations}\label{eq:generator.model}
\begin{align}
	\ddt \theta_\g&=\omega_\g,\\
	J_\g \omega_\g \ddt \omega_\g&= P_\g -b_\g \sin(\theta_\g-\theta_\gsc),\\
	T_\g \ddt P_\g&=-P_\g-k_\g \omega_\g.
\end{align}
\end{subequations}
with inertia constant $J_\g \in \mathbb{R}_{>0}$, turbine/governor time constant $T_\g \in \mathbb{R}_{>0}$, governor gain $k_\g \in \mathbb{R}_{>0}$, deviation of the turbine power from its setpoint $P_{\delta,\g}  \in \mathbb{R}$, deviation of the generator frequency from the nominal frequency $\omega_{\delta,\g} \in \mathbb{R}$, and voltage phase angle $\theta_{\delta,\g}$ relative to a reference frame rotating at the nominal frequency $\omega_0$. Next, we assume that AC voltage magnitudes are constant and $T^\gsc_\dc=T^\msc_\dc=0$. Combining \eqref{eq:machine.model}, \eqref{eq:converter.model}, \eqref{eq:powerflow}, \eqref{eq:GSC voltage angle}, \eqref{eq:MSC voltage frequency},  \eqref{eq:WTpowerlin}, \eqref{eq:generator.model}, and linearizing at zero phase angle difference, results in the small signal model
\begin{align} \label{eq:dynamics.compact}
 \!\!T  \ddt \! \begin{bmatrix}
		\varrho_\delta \\ \omega_\delta \\ v_{\delta,\dc} \\ P_{\delta,\g} \end{bmatrix} \!\!=\!\!  \begin{bmatrix} -K^\prime_d \mathbbl{1}_{2\times2} \mc{B} & \!-I_2 & \!K_\theta \mathbbl{1}_2 & \!\mathbbl{0}_{2} \\
		\mc B & \!-K_\wt & \!\mathbbl{0}_{2}&  \!\mathbbl{e}_2\\
		-\mathbbl{1}^\top_2 \mc B & \!0 & \!0  & \!0\\
		0& \!-k_\g & \!\mathbbl{0}_{2}^\top & \!-1 \end{bmatrix} \!\! \begin{bmatrix}
		\varrho_\delta \\ \omega_\delta \\ v_{\delta,\dc} \\ P_{\delta,\g} \end{bmatrix} \!
\end{align}
where ${\varrho_\delta} \coloneqq (\theta_\gsc-\theta_\g , \theta_\msc-\theta_r) \in \mathbb{R}^2$, $\omega_\delta \coloneqq (\omega_{\delta,\g}, \omega_{\delta,r})\in \mathbb{R}^2$, $\omega_{\delta,r}\coloneqq\omega_r-\omega^\del_r$ denote phase angle differences between the GSC and SG as well as the MSC and PMSG, and the SG and PMSG frequency deviations. Moreover, we use $\mathbbl{e}_2=(1,0) \in \mathbb{R}^2$, $\mathbbl{1}_2=(1,1) \in \mathbb{R}^2$, and $\mathbbl{1}_{2\times2}=\mathbbl{1}_2\mathbbl{1}_2^\top$. Finally, $\mc B \coloneqq \diag(b_\g,b_\msc)$, {$K^\prime_d\coloneqq \diag(K_d^\gsc/C_\dc, K_d^\msc/C_\dc)$}, {$K_\theta\coloneqq\diag(K_{\theta}^\gsc, K_{\theta}^\msc)$}, $K_\wt\coloneqq\diag(0,K_{\omega_r}+K_\beta K_p)$, $J\coloneqq\diag\!(J_\g \omega_0,J_\wt \omega^\del_r\!)$ and $T\coloneqq \diag(I_2,J,C_\dc,T_\g)$ denote the susecptance matrix, control gains, wind turbine sensitivity, and matrix of inertia and time constants.

\begin{theorem}{\bf{(Frequency and DC voltage stability)}}\label{th:stab}
 Assume that $K_{\omega_r}+K_\beta K_p \geq 0$ and $K_d^\gsc / K_\theta^\gsc = K_d^\msc / K_{\theta}^\msc$, then the system \eqref{eq:dynamics.compact} is asymptotically stable with respect to the origin.
\end{theorem}
\begin{IEEEproof}
Let $x\coloneqq(\mc B \varrho_\delta,\omega_{\delta,r}, v_{\delta,\dc}, P_{\delta,\g}) \in \mathbb{R}^6$ and consider the LaSalle function $V = x^\top \mc M x$ with
\begin{align*}
 \mc M \coloneqq \tfrac{1}{2} \diag\left({(K_\theta \mc B)^{-1}},K_\theta^{-1} J, C_\dc, T_\g / (k_\theta^\gsc k_\g)\right).
\end{align*}
It can be verified that $\ddt V=-x^\top \mc V x \leq 0$, where
	\begin{align*}
		\mc V = \diag\left(\tfrac{K_d^\gsc}{K_\theta^\gsc C_\dc}\mathbbl{1}_2\mathbbl{1}_2^\top , K_\wt / K^\msc_\theta , 0, (K_\theta^\gsc K_\g)^{-1}\right).
	\end{align*}
Next, for $K_\wt \in \mathbb{R}_{\geq 0}$, the set $\mc E \coloneqq \{ x \in \mathbb{R}^6 \vert \ddt V =0\}$ can be rewritten as {$\mc E=\{ x \in \mathbb{R}^6 \vert \mathbbl{1}_2 \mathbbl{1}_2^\top \mc B \varrho_\delta = \mathbbl{0}_2, K_\wt/K_\theta^\msc \omega =\mathbbl{0}_2, P_{\delta,\g}=0\}$}. On the maximal invariant set $\mc S \subseteq \mc E$ contained in $\mc E$, it needs to hold that $\frac{\diff^k}{\diff t^k} \mathbbl{1}_2 \mathbbl{1}_2^\top \mc B \varrho_\delta = \mathbbl{0}_2$, {$K_\wt/ K_\theta^\msc \frac{\diff^k}{\diff t^k} \omega =\mathbbl{0}_2$} and $\frac{\diff^k}{\diff t^k} P_{\delta,\g}=0$ for $k \in \{1,\ldots,3\}$ and the dynamics \eqref{eq:dynamics.compact}. It can be verified that this invariance conditions holds if and only if $x=\mathbbl{0}_6$, i.e., $\mc S = \{\mathbbl{0}_6\}$. Next, we note that $\mc M$ is positive definite and $\mc V$ is positive semidefinite, i.e., all sublevel sets of $V$ and trajectories of \eqref{eq:dynamics.compact} are bounded. Using LaSalle's invariance principle and linearity of the system \eqref{eq:dynamics.compact}, it follows that the maximal invariant set $\mc S = \{\mathbbl{0}_6\}$ contained in $\ddt V=0$ is asymptotically stable. 
\end{IEEEproof}

This result highlights that the AC voltage phase angles and frequencies in the AC system and on the turbine side of the MSC synchronize, and frequency and DC-link voltage stability are ensured by the primary frequency control response of the synchronous generator's turbine/governor system and response of the wind turbine to a change in rotor speed or pitch angle.

\subsection{Steady-state control specifications and control gains}\label{sec:gainsselection}
At steady state, the GSC and MSC frequency deviations prescribed by \eqref{eq:GSC voltage angle} and \eqref{eq:MSC voltage frequency}  are proportional to DC-link voltage deviation, i.e., 
\begin{subequations}\label{eq:Vdc droop}
 \begin{align} 
\omega_\gsc-\omega_0&=K^\gsc_\theta (v_\dc-v_\dc^\star), \label{eq:GSC Vdc droop}\\
\omega_\msc-\omega^\del_r&=K^\msc_\theta (v_\dc-v_\dc^\star). \label{eq:MSC Vdc droop}
\end{align}
\end{subequations}

Moreover, as shown in the previous section, the MSC and PMSG frequencies will synchronize through the power flow \eqref{eq:mscpower}, i.e., $\omega_\msc = \omega_r$, and the GSC frequency is synchronous with the grid frequency, i.e., $\omega_\gsc = \omega_\g$. Together, with \eqref{eq: WT power combine rotor & pitch}, \eqref{eq:Vdc droop}, and $\omega_\msc = \omega_r$ and linearizing around the nominal operating point results in the steady-state response of the wind turbine active power to GSC frequency deviations
\begin{align} \label{eq: WT power droop with rotor & pitch}
P_\wt=P^\del_\wt-\frac{K^\msc_\theta}{K^\gsc_\theta}(K_{\omega_r}+K_\beta K_p)(\omega_\g-\omega_0).
\end{align}
In other words, the wind turbine provides a primary frequency control response with droop coefficient
\begin{align} \label{eq: WT droop coefficient}
m_p=\frac{K^\gsc_\theta}{K^\msc_\theta(K_{\omega_r}+K_\beta K_p)}.
\end{align}
When selecting the control gains $K^\gsc_\theta\in\mathbb{R}_{>0}$ and $K^\msc_\theta\in\mathbb{R}_{>0}$, the resulting droop coefficient $m_p$ as well as the operational limits of the DC-link voltage, wind turbine speed, and pitch angle have to be considered.

In particular, we use $\Delta \omega^{\max}_\g$ to denote the largest expected grid frequency deviation (e.g., the boundary of the normal operating range) and $\Delta v^{\max}_\dc$ to denote the largest acceptable DC-link voltage deviation. Then, considering \eqref{eq:GSC Vdc droop},
\begin{align}\label{eq:vdc limit}
 K^\gsc_\theta \leq \frac{\Delta \omega^{\max}_\g}{\Delta v^{\max}_\dc}
\end{align}
needs to hold to ensure that the DC-link voltage remains within the acceptable range in steady-state. Moreover, decelerating the rotor to $\omega_r < \omega_{r \text{mpp}}$ results in $K_{\omega_r}<0$ and small signal instability of the proposed control and standard controls (see e.g., \cite{janssens2007active}). Therefore, the mapping between system frequency deviation and rotor speed deviation needs to be restricted. Substituting \eqref{eq:MSC Vdc droop} into \eqref{eq:GSC Vdc droop}, it can be seen that 
\begin{align} \label{eq: Kt limit}
\frac{K^\msc_\theta}{K^\gsc_\theta} \leq \frac{\omega^\del_r-\omega^\text{mpp}_r}{\Delta \omega^{\max}_\g}
\end{align}
needs to hold to ensure that $\omega_r = \tfrac{K^\msc_\theta}{K^\gsc_\theta} \omega_g \geq \omega^\mpp_r$ in steady-state. Finally, considering \eqref{eq:beta_ref},  
\begin{align} \label{eq: Kp limit}
K_p \leq \frac{K^\gsc_\theta}{K^\msc_\theta} \frac{\beta_\del}{\Delta \omega_{g\text{max}}}.
\end{align}
needs to hold to ensure that $\beta_\refe \geq 0$ in steady-state and the primary frequency response with gain $m_p$ given by \eqref{eq: WT droop coefficient} is provided across the specified range of grid frequencies. 

We emphasize that the constraints \eqref{eq:vdc limit} to \eqref{eq: Kp limit} are fundamental restrictions of the wind turbine system that affect all possible controls. One of the key features of the proposed dual-port GFM control is that it makes the constraints explicit by directly mapping imbalances between the power system, DC-link, and wind turbine rotor. In contrast, using standard controls the requirements on the wind turbine and DC-link are obscured by the assumption that the DC-link voltage is regulated to its nominal value either through the GSC or MSC.

\begin{figure}
\centering
\includegraphics[width=8cm]{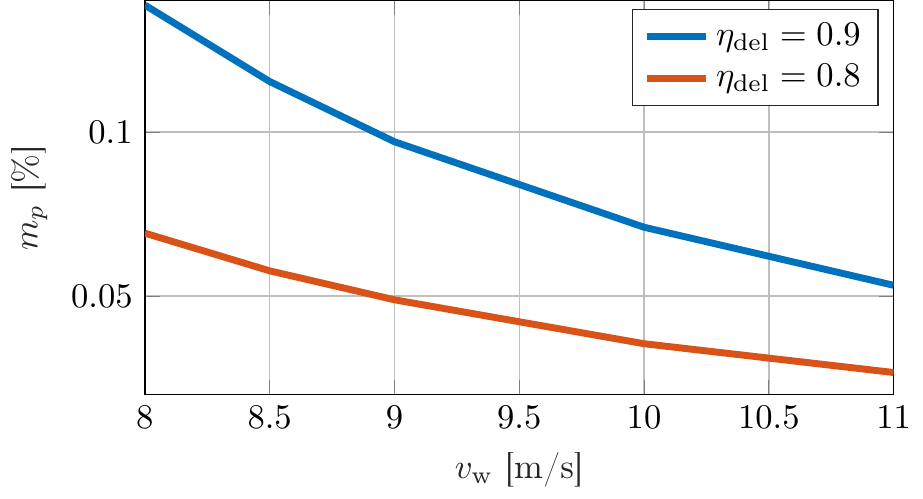}
\caption{Smallest achievable frequency droop percentage for a $5~\mathrm{MW}$ wind turbine \cite{NREL5MW}, the curtailment strategy proposed in Sec.~\ref{sec:curtailment}, and control gains according to Sec.~\ref{sec:gainsselection}.\label{fig:smallestmp}}
\end{figure}

Finally, the smallest achievable droop percentage for a $5~\mathrm{MW}$ wind turbine \cite{NREL5MW} is shown in  Fig. \ref{fig:smallestmp} for $\Delta \omega^{\max}_\g = 0.005~\mathrm{pu}$, the curtailment strategy discussed in Sec.~\ref{sec:curtailment}, and selecting the largest gain allowed by \eqref{eq: Kp limit} for the largest $K^\msc_\theta$ allowed by \eqref{eq: Kt limit}. In other words, giving priority to rotor speed based control to maximize the inertia contribution of the wind turbine. It can be seen that the steady-state constraints  \eqref{eq:vdc limit} to \eqref{eq: Kp limit} restrict the ability of the $5~\mathrm{MW}$ wind turbine used for this example \cite{NREL5MW} to provide stiff primary frequency control at low wind speeds and low curtailment. Nonetheless, the primary frequency response capabilities of the wind turbine are significant and typical droop percentages of conventional generation can be achieved at higher wind speeds and/or with increased curtailment.

\section{Numerical Case Study}
To illustrate and validate the performance of the proposed control a case study is used to compare the response of standard grid-following MPPT control (GFL MPPT), dual-port GFM control without reserves (GFM MPPT), and dual-port GFM with reserves and frequency response (GFM FR). 

\subsection{Test system, operating point, and control gains}
\begin{figure}[t!!]
\centering
\includegraphics[width=8.5cm]{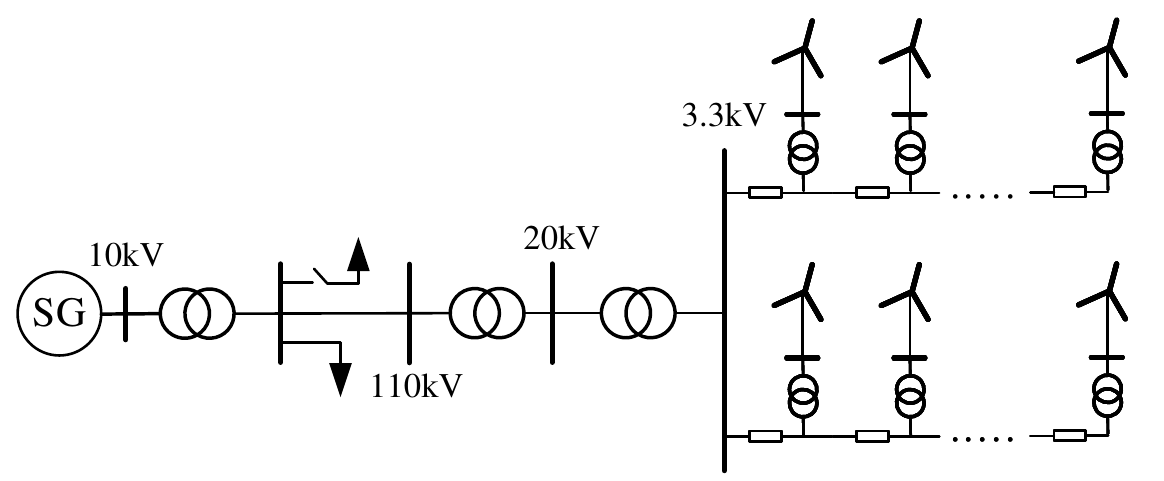}
\caption{Test system with a $210~\mathrm{MW}$ synchronous generator, ten $5~\mathrm{MW}$ wind turbines, and constant power loads.}\label{fig:test system}
\end{figure}

To illustrate and validate the performance of the proposed control, we consider the test system depicted in Fig.~\ref{fig:test system} that contains a $210~\mathrm{MVA}$ synchronous generator (SG) and an aggregate model of a PMSG wind turbines with back-to-back two-level voltage source converters representing ten identical $5~\mathrm{MW}$ wind turbines. The synchronous generator is equipped with a turbine/governor system with $5\%$ droop and  automatic voltage regulator. The parameter of the PMSG wind turbines and two-level voltage sources converters are summarized in Table~\ref{tab:WT parameter}. The system base load is modeled using a $100~\mathrm{MW}+j5~\mathrm{MVar}$ constant power load and an additional $20~\mathrm{MW}$ constant power load is used to simulate a load step. Moreover, we consider three wind speeds for which the curtailment strategy proposed in Sec.~\ref{sec:curtailment} results in curtailment (i) using only rotor speed, (2) using rotor speed and pitch angle, and (3) using only pitch angle. In all three scenarios, the wind turbines operate at $10\%$ curtailment (i.e., $\eta_\del=0.9$). The resulting control parameters are summarized in Table~\ref{tab:ctrparind} and Table~\ref{tab:ctrpar}. We emphasize that, in all scenarios $K_d^\msc = K_d^\gsc K_{\theta}^\msc / K_\theta^\gsc$ (see Theorem \ref{th:stab}), and, in GFM MPPT mode, the gain $K_\theta^\msc=K_\theta^\gsc$ is used.
\begin{table}[h!!!] 
\caption{Wind turbine parameters}
\label{tab:WT parameter}
\centering
\begin{tabular}{|c|c|c|}
\hline
Component & Parameter & Value\\
\hline
\multirow{3}*{WT} & Nominal power & $5~\mathrm{MW}$ \\
\cline{2-3}
~ & Nominal rotor speed & $1.37~\mathrm{rad/s}$\\
\cline{2-3}
~ & Total moment of inertia & $35.328 \times {10^6}^{\vphantom{A}} \mathrm{kg~m^2}$\\
\hline
\multirow{2}*{DC-link} & Voltage & $7.92~\mathrm{kV}$\\
\cline{2-3}
~& Capacitance & $31.88~\mathrm{mF}$\\
\hline
\multirow{3}* {GSC filter} & Resistance & $0.0218~\Omega$\\
\cline{2-3}
~& Inductance & $0.693~\mathrm{mF}$\\
\cline{2-3}
~& Capacitance & $73 ~\mathrm{\mu F}$\\
\hline
\end{tabular}
%
\vspace{1em}
\caption{Control Parameters (independent of $v_\w$)} \label{tab:ctrparind}
\centering
\addtolength{\tabcolsep}{-4pt}    
\renewcommand{\arraystretch}{1.2}
\begin{tabular}{|c|c|c|c|c|c|c|}
\hline
$K^\gsc_\theta$ & $K^\gsc_d$ & $T^\gsc_\dc$,$T^\msc_\dc$ & $T^\gsc_v$,$T^\msc_v$ & $K^\gsc_q$ & $K^\msc_q$ & $K^\msc_\theta$ (MPPT)\\
\hline
$0.5$ & $0.0067~\mathrm{pu}$ & $0.05~\mathrm{s}$ & $0.05~\mathrm{s}$ & $0.02~\mathrm{pu}$ & $0.05~\mathrm{pu}$ & $0.5~\mathrm{pu}$\\
\hline
\end{tabular}
\vspace{1em}
\caption{Control Parameters in GFM FR Control Mode (rounded) \label{tab:ctrpar}}
\centering
\addtolength{\tabcolsep}{-0pt}    
\begin{tabular}{|c|c|c|c|c|c|c|c|}
\hline
$v_\w$ & $\omega^\del_r$ & $\beta_\del$ & $K^\msc_\theta$ & $K_p$ & $K_{\omega_r}$  & $K_\beta$ & $m_p$\\
\hline
$8~\mathrm{m/s}$ & $1.16~\mathrm{pu}$ & $0^{\circ}$ & $15.1~\mathrm{pu}$ & $0~\tfrac{\circ}{\mathrm{pu}}$ & $0.119~\mathrm{pu}$ & $0~\tfrac{\mathrm{pu}}{\circ}$ & $27.7\%$\\
\hline
$10~\mathrm{m/s}$ & $1.2~\mathrm{pu}$ & $3^{\circ}$ & $6.6~\mathrm{pu}$ & $22.7~\tfrac{\circ}{\mathrm{pu}}$ & $0.082~\mathrm{pu}$ & $0.02~\tfrac{\mathrm{pu}}{\circ}$ & $14.2\%$\\
\hline
$12~\mathrm{m/s}$ & $1.2~\mathrm{pu}$ & $5.4^{\circ}$ & $1~\mathrm{pu}$ & $270~\tfrac{\circ}{\mathrm{pu}}$ & $0~\mathrm{pu}$ & $0.083~\tfrac{\mathrm{pu}}{\circ}$ & $2.3\%$\\
\hline
\end{tabular}
\end{table} 

\subsection{Low wind speed}
\begin{figure}[b!!]
\centering
\includegraphics[width=9cm]{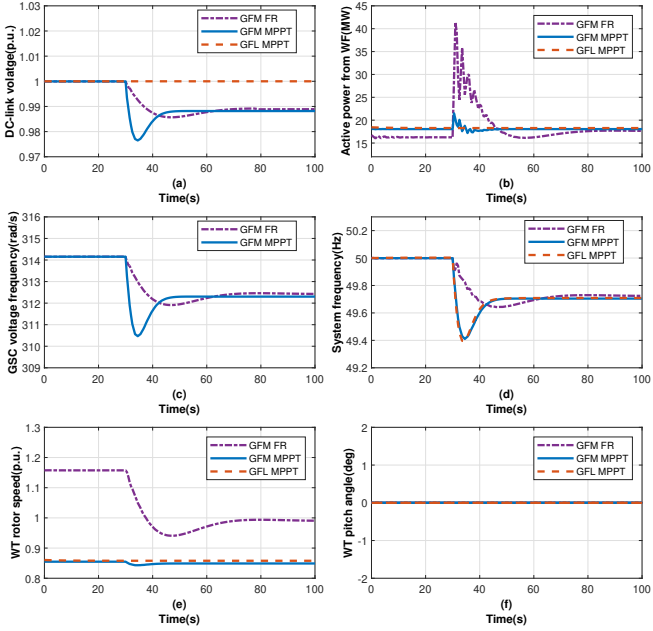}
\caption{Simulation results at $v_\w=8~\mathrm{m/s}$.}\label{fig:8m/s}
\end{figure}

We first consider an operating point with low wind speed ($v_\w=8~ \mathrm{m/s}$). Simulation results for a $20~\mathrm{MW}$ load step at $t=30~\mathrm{s}$ are depicted in Fig.~\ref{fig:8m/s}. Using standard GFL MPPT control, the DC voltage is maintained at $1~\mathrm{pu}$ throughout simulation and the wind turbines do not provide any grid support. In contrast, using the proposed control approximately at the MPP, the DC voltage deviation is approximately proportional to the system frequency deviation when  the load step occurs (see Fig.~\ref{fig:8m/s}(a)). Moreover, as predicted in Sec.~\ref{sec:analysis}, the GSC frequency synchronizes to the system frequency (see Fig.~\ref{fig:8m/s} (c) and (d)) illustrating that grid synchronization can be achieved without relying on a PLL. However, if the wind turbine operates approximately at the MPP and with low gain $K^\msc_\theta$, a small inertia response and no significant frequency response are provided. In GFM FR mode the wind turbine operates at $10\%$ curtailment (i.e., $\eta_\del=0.9$ achieved only through increased rotor speed (see Sec.~\ref{sec:curtailment}) and the control gains are selected using the steady-state specifications in Sec.~\ref{sec:gainsselection}. In this case, the rotor speed reduces following the load step, kinetic energy from the wind turbine rotor is released to the grid, and the wind power generation increases (see Fig. ~\ref{fig:8m/s} (b)). As shown in Fig.~\ref{fig:8m/s} (c), the frequency response of the overall system is improved significantly by the wind turbines response, i.e., the rate of change of frequency (RoCoF) and frequency nadir are significantly improved. Compared with both GFL MPPT and GFM MPPT control, the nadir frequency increases from 49.397 $\mathrm{Hz}$ to 49.644 $\mathrm{Hz}$, and the steady state frequency increases from $49.709~\mathrm{Hz}$ to $49.724~\mathrm{Hz}$. 

\subsection{Medium wind speed}
\begin{figure}[b!!]
\centering
\includegraphics[width=9cm]{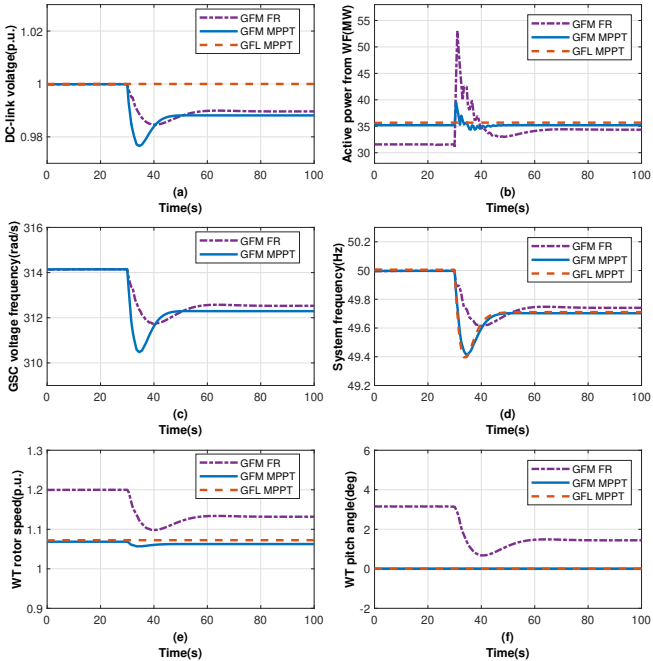}
\caption{Simulation results at $v_\w=10~\mathrm{m/s}$.}\label{fig:10m/s}
\end{figure}

Simulation results for medium wind speed (i.e., $v_\w=10~\mathrm{m/s}$) are shown in Fig.~\ref{fig:10m/s}. Similarly to the low wind speed scenario in the previous section, the wind turbines cannot provide a frequency response using the GFL MPPT strategy. Moreover, using the proposed dual-port GFM control, the DC voltage deviation again reflects the system imbalance and grid synchronization is successfully achieved without PLL as shown in Fig.~\ref{fig:10m/s} (c) and (d). The curtailment strategy in Sec.~\ref{sec:curtailment} and $10\%$ curtailment (i.e., $\eta_\del=0.9$) result in both curtailment through increased rotor speed and pitch angle. Using dual-port GFM control at the resulting operating point, kinetic energy is released to the system as the DC voltage and rotor speed decrease after the load step. This results in a significant improvement of the the frequency nadir and RoCoF compared to GFL MPPT control and GFM MPPT control. However, compared with the low wind speed scenario, the difference between $\omega^\mpp_r$ and $\omega^\del_r$ is smaller and less kinetic energy is available for grid support in the medium wind speed scenario and the frequency nadir is slightly lower ($49.613~\mathrm{Hz}$). However, larger power reserves are obtained resulting in a significant increase of the wind turbine's frequency droop coefficient. Specifically, the pitch angle decreases as the rotor speed decreases and more wind power is captured and injected into the system (see Fig.~\ref{fig:10m/s} (b)). This results in an improved steady state system frequency (see Fig.~\ref{fig:10m/s} (d)). 

\subsection{Wind speed above rated speed}
\begin{figure}[h!!]
\centering
\includegraphics[width=9cm]{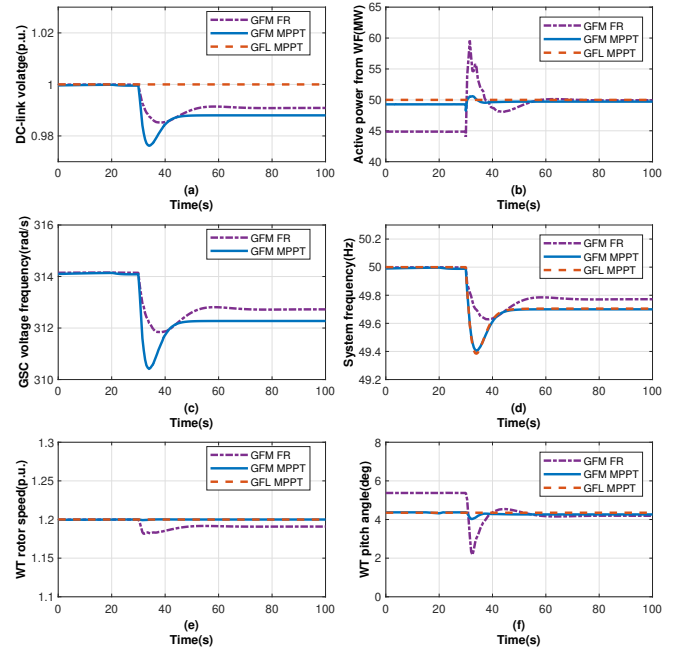}
\caption{Simulation results at $v_\w=12~\mathrm{m/s}$.}\label{fig:12m/s}
\end{figure}

Finally, simulation results for a wind speed (i.e., $v_\w=12 \mathrm{m/s}$) beyond the rated wind speed (i.e., $v_w=11.23 \mathrm{m/s}$). In this scenario, the rotor speed reaches its maximum, the pitch angle is $4.3^{\circ}$ for MPPT control mode (see Fig.~\ref{fig:12m/s} (e) and (f)), and $10\%$ curtailment (i.e., $\eta_\del=0.9$) can only be achieved through increasing the pitch angle to $\beta_\del=5.4^{\circ}$. As shown in Fig.~\ref{fig:smallestmp} a significant droop response can be achieved at high wind speeds and $K_\theta^\msc=1$ has been selected to maintain a frequency droop coefficient above $2\%$. Consequently, GFM FR control results in a DC voltage and rotor speed decrease after the load step  (see Fig.~\ref{fig:12m/s} (e)), the pitch angle decreases (see Fig. \ref{fig:12m/s}(f)) and wind power generation increases. Compared with MPPT control, the frequency nadir, RoCoF, and steady-state frequency significantly improve.
\section{Conclusion and Outlook}
In this work, a novel dual-port grid-forming controller for wind turbines using a permanent magnet synchronous generator interfaced through back-to-back voltage source converters is presented. The proposed controller unifies standard functions of grid-following control (i.e., MPPT) and grid-forming controls (i.e., primary frequency control) in a single simple controller. In contrast to standard grid following control, the proposed grid side converter control imposes a well defined AC voltage waveform on the grid and achieves synchronization and DC voltage control without relying on a PLL. In contrast to standard grid-forming control, the machine side converter control supports both operation at the MPP and curtailed operating points. Together with a suitable curtailment strategy, the dual-port grid-forming control utilizes both rotor speed and pitch angle control to provide inertia support and fast frequency response. Dynamic stability and the steady-state response are analyzed and a comprehensive case study is used to illustrate and validate the results. Studying trade offs between the mechanical stress on the wind turbines, size of the DC-link capacitor, and achievable grid-support is seen as an interesting topic for future work.

\bibliographystyle{IEEEtran}
\bibliography{IEEEabrv,irep_2022}
\end{document}